# A New Ultrasonic Transducer Sample Cell for In-Situ Small-Angle Scattering Experiments

Sudipta Gupta[1], Markus Bleuel[2,3] and Gerald J. Schneider[1]

[1]Department of Chemistry and Department of Physics, Louisiana State University, Baton Rouge, LA 70803, United States

[2]NIST Center for Neutron Research, National Institute of Standards and Technology, Gaithersburg, MD 20899, United States

[3]Department of Materials Science and Engineering, University of Maryland, College Park, MD 20742-2115, United States

**ABSTRACT**

Ultrasound irradiation is a commonly used technique for nondestructive diagnostics or targeted destruction. We report on a new versatile sonication device that fits in a variety of standard sample environments for neutron and X-ray scattering instruments. A piezoelectric transducer permits measuring of the time-dependent response of the sample *in situ* during or after sonication. We use small-angle neutron scattering (SANS) to demonstrate the effect of a time-dependent perturbation on the structure factor of micelles formed from sodium dodecyl sulfate (SDS) surfactant molecules. We observe a significant change during and after sonication in the micellar relaxation after interruption induced by ultrasound irradiation. We also observe a time-dependent relaxation to the equilibrium values of the unperturbed system. The strength of the perturbation of the structure factor depends systematically on the duration of sonication. The relaxation behavior can be well reproduced after multiple times of sonication. Accumulation of the recorded intensities of the different sonication cycles improves the signal-to-noise ratio and permits reaching very short relaxation times. In addition, we present SANS data for the micellar form factor on alkyl-polyethylene-oxide (PEO) surfactant molecules irradiated by ultrasound. Due to the flexibility of our new *in-situ* sonication device, different experiments can be performed, e.g. to explore molecular potentials in more detail by introducing a systematic time-dependent perturbation.

**I. INTRODUCTION**

Irradiating biological and soft matter samples by ultrasound and high frequency acoustic instrumentation is gaining more and more importance.[1-8] On the one hand, ultrasound imaging is a very important non-invasive technology that can image materials at μm length-scale.[9, 10] On the other hand, high intensity focused ultrasound has found application in tuning the shape of biodegradable polymers,[11] rupturing lipid coated micro-bubbles,[12] and crossing epidermis (skin)[13] for drug delivery. Low intensity ultrasound at a frequency of 1 – 2 MHz, is used in sonodynamic therapy for treating cancer.[14] Despite the strong influence that ultrasound may have on the materials, its time dependent effect on the nm length-scale has not yet been explored.

Scattering experiments are well established in measuring the structure and dynamics of soft matter samples, such as polymers,[15-19] polymer composites and nanocomposites[20-27] and polymer aggregates like micelles,[28-30] biopolymers,[31] proteins,[32, 33] and glasses on the nm length scale.[34-36] A variety of methods to manipulate samples and to record in parallel the morphology or dynamics is described in the literature.[37-39] For example, sample environments to accurately regulate the physical parameters, like temperature, magnetic field, pressure or humidity belong to the standard equipment of SANS instruments. While these techniques have its primary focus on keeping a certain parameter fixed, other techniques such as *in situ* stretching or rheology experiments manipulate the mesostructure of the sample during measurement.[40-42]

Recording SANS data during static deformation or low frequency oscillatory shear rheology enable us to measure the influence of an external force on the structure at the nm length scale and thereby to understand the molecular interaction in details. [40, 41] Exploring details of such interactions allow us to derive material models that can themselves be used to understand the macroscopic properties. [43-47]

A high frequency perturbation of samples is possible by exposure to an ultrasonic field. We designed and built a new sample cell by incorporating an ultrasonic transducer, that was used to successfully record scattering diagrams *in situ*, during and immediately after sonication. It has been well documented in the literature that small-angle neutron scattering (SANS) experiments on micelles show a clear response to external perturbations.[28, 48-50] Therefore, for our first experiments, we have chosen to use the sodium dodecyl sulfate (SDS) surfactant micelles in aqueous solution, whose unperturbed state is well-established. Following Bergström and Pedersen,[51] SDS forms ellipsoidal micelles at 40°C and a concentration, $\phi_w$ = 0.5%. Form factor analysis yields a semi-major axis of 23 Å and semi-minor axis of 13 Å. The number of surfactant molecules per micelle are defined by their aggregation number, $N_{agg}$ = 54. Following a detailed SANS investigation, the aggregation number of SDS micelles was found to decrease with increasing temperature and decreasing concentration.[52] At 25°C and, $\phi_w$ = 5%, a pronounced structure factor shows up in the scattering intensity, where, $N_{agg}$ = 89 and micellar volume of 3.64 ×10$^4$ Å$^3$ was reported.[52] We will use this particular concentration and temperature for our investigation to determine the time constant associated with micellar self-assembly followed by ultrasound induced disintegration at a fixed temperature. During both disintegration and self-assembly we observe a time dependent change, showing the importance of our new *in situ* sonication tool.

In addition to SDS we present the in-situ SANS diffraction data for commercial (Brij100) alkyl-poly (ethylene oxide) (n-PEO) micelles in $D_2O$. We have chosen n-PEO surfactant since it is a well-established model system for frozen micelles.[53, 54]

## II. DESIGN OF THE ULTRASONIC TRANSDUCER CELL

In this section, some technical details about the novel ultrasonic sample cell are given. To make the cell user friendly two important design criteria are satisfied. First, the setup fits inside the standard sample environments so that we have access to a wide range of well-established instrument controls like variation of temperature, humidity or magnetic fields as if required by the user community. Second, the options to vary sample thickness to fit the needs of different soft matter samples. Figure 1(a) illustrates the schematic representation of the ultrasonic transducer cell. The outer dimensions are about 25 × 25 × 50 mm. Cylindrical passages for the neutron beam (Ø about 18 mm) are drilled on each side and threaded to fit aluminum (Al) caps. The distance between the caps is variable and defines the sample thickness in the neutron beam. We used 2 mm path length for our SANS experiment. The transducer is glued with epoxy onto one of the aluminum caps as indicated. If needed, a second transducer can be added. The bottom opening was sealed with a screw coated with Teflon ribbon and allows for future upgrades of the cell like a sensor to measure the sonic field. As shown in Figure 1(b), the transducer is a piezoelectric ring, purchased from Steminc (part number: SMR1585T07111R), with an outer diameter of 15 mm, a thickness of 0.7 mm and inner diameter of 8.5 mm. The ring transducer is essentially a ceramic ring (piezo material SM111) with silver (Ag) electrodes on the same side. It resonates at thickness mode with a resonant frequency of 2.95 ± 0.09 MHz, with electromechanical coupling coefficient ≥ 49%, it has a resonant impedance ≤ 0.65Ω, and a static capacitance of 1840 ± 276 pF at 1 KHz frequency. The glued transducer is in direct contact with the sample in the sample chamber (Figure 1). The neutron beam is guided through the central passage of the transducer and is limited in diameter using a standard ¼ inch (6.35 mm) SANS Cadmium aperture. The only material in the neutron beam besides the sample are the aluminum end caps with a total wall thickness of less than 1mm. If required Al can be replaced by other materials like niobium, quartz or sapphire. More than 95% empty cell transmission was measured for neutrons.

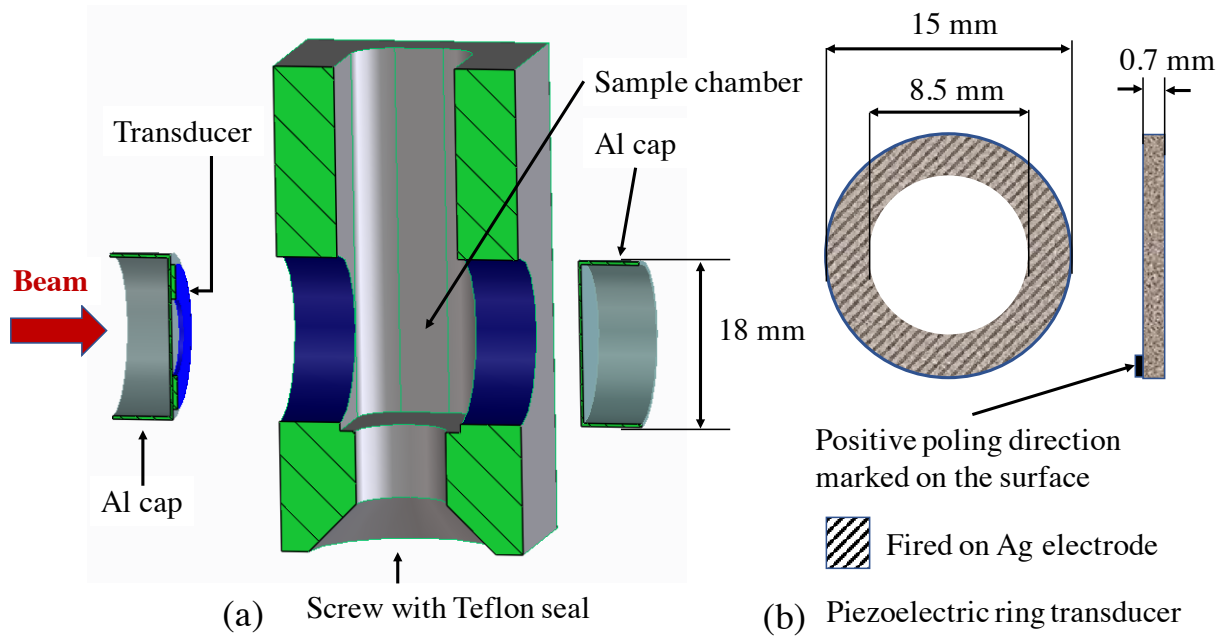

Fig.1. (1 column, color) (a) *Schematic representation of the Ultrasonic transducer cell, (b) Schematics of piezoelectric ring transducer from Steminc (top and side view).*

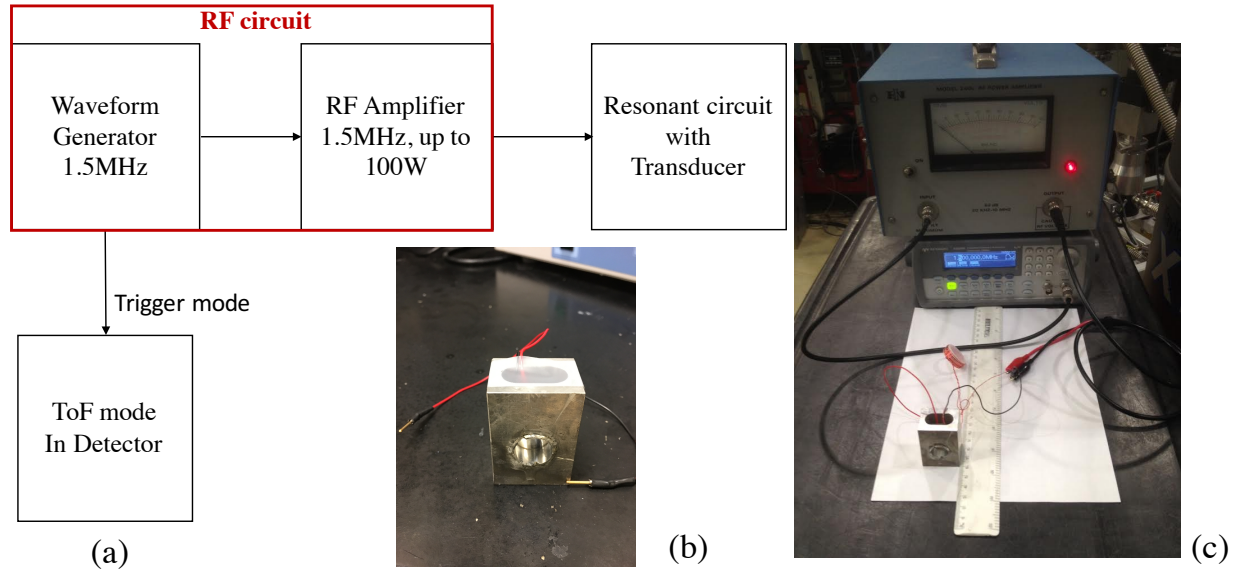

Fig.2. (1 column, color) (a) *Schematic block diagram of the circuit used in this experiment. (b) Photo of the ultrasonic transducer cell (c) Photo of the RF circuit and the cell.*

Figure 2(a) illustrates the block diagram of the circuit along with (b) the photo of the transducer cell and (c) that of the circuit. The radio frequency (RF) circuit consists of a RF amplifier which is driven by a waveform generator producing a continuous sinusoidal waveform of fixed frequency and can feed a power up to 100 W. By connecting the RF circuit to a serial resonant circuit, we can easily vary the frequency up to 3 MHz by changing the inductance. For our tests, we used an air coil ($N = 10$, length 5 mm, $\varnothing = 25$ mm, see Figure 2(c)). We have also tested the option to generate ultrasonic signal at a fixed frequency by connecting the resonant circuit of the transducer to a separately purchased mist generation kit (SMUTK2500RS112) which is self-tuning to a frequency predefined by the transducer and an onboard inductance. As Figure 2(a) indicates, the option to use the setup to trigger the SANS detector in pulsed mode for time-of-flight (ToF) experiments is available without any changes to the electronics. A transistor-transistor-logic (TTL) signal produced by the waveform generator can be used to trigger the reset of the timing mechanism for the detector at a fixed interval which synchronizes the neutron measurement with the ultrasonic set-up.

A typical frequency range of 1 to 2 MHz is used for ultrasound in sonodynamic therapy.[14] So in our tests we used the ultrasonic transducer at 1.5 MHz, feeding a power of 10 – 30 W to the resonant circuit yielding an estimated ultrasonic field intensity of, $I_0 \sim 10$ W·cm$^{-2}$ at 100% pulse amplitude. In addition, such a cell can be used for ultrasonic imaging and diagnostic echocardiography using a large diameter transducer with higher frequency and a receiver. It will open up the possibilities to acquire ultrasonic images and scattering data of different soft matter samples simultaneously.

**III. SANS EXPERIMENTAL DETAILS**

The SANS experiments were carried out using the NGB 30 m SANS instrument of the NIST Center for Neutron Research (NCNR) at National Institute of Standards and Technology (NIST).[55] The sample-to-detector distance was fixed to 2 m and the neutron wavelength was, $\lambda$ = 6 Å. This configuration covers a $Q$ - range from $\sim$ 0.02 Å$^{-1}$ to $\sim$ 0.23 Å$^{-1}$, where, $Q = 4\pi\sin(\theta/2)/\lambda$, for the scattering angle $\theta$. A wavelength resolution of, $\Delta\lambda/\lambda$ = 10%, was used. All data reduction into intensity $I(Q)$ vs. momentum transfer $Q = |\vec{Q}|$ was carried out following the standard procedures that are implemented in the NCNR macros for the Igor software package.[56] The data are scaled into absolute units (cm$^{-1}$) using a direct beam, and a detector sensitivity correction was done with a plexiglass measurement. The solvents and empty cell are measured separately as backgrounds. The ultrasonic transducer cell was mounted inside a standard multi-position heating/cooling block for SANS designed to hold up to 9 standard demountable titanium cell holders. The block has the dimension of 53 × 35 × 38.1 mm, with a beam aperture of 12.5 mm. The heating element in the block provides a temperature range from ambient to 300°C, with an accuracy of ±0.5°C. Quartz or silicon window are used to prevent heat loss by convection.

## IV. RESULTS

**SDS micelles**

Figure 3 displays the SANS intensity, $I$, as a function of the momentum transfer, $Q$, for, $\phi_w = 5\%$, SDS in $D_2O$. The data is presented over a $Q$- range centering around the structure factor peak, $S(Q)$, in presence and absence of ultrasonic pulses for several selected times, $t$. The measurements were performed at a constant temperature of 25°C, following a three-step protocol. At first, we measured unperturbed sample to determine the initial state at time, $t = 0$ s. Second, the sample was sonicated for $t_p = 65$ s (pulse on), at a frequency of, $\nu_s = 1.5$ MHz, and amplitude 100% that corresponds to an intensity, $I_0 \sim 10$ W·cm$^{-2}$. Third, we waited for, $t_{off} = 1305$ s (pulse off). Hereafter, we use the notion "pulse" for the perturbation by ultrasound of a certain amplitude and duration at a constant frequency $\nu_s$. The scattering diagrams were recorded with a repetition rate of 30 s for a total period of 1335 s (off-on-off cycle). Figure 3(a) reports the SANS data acquired during ultrasonic irradiation time. Here, $t = 0$ s represent the unperturbed sample. Figure 3(b) shows SANS data after the sonication was switched off. The legends in Figure 3(b) indicates the SANS scattering pattern evolving with time once the ultrasonic pulse was turned off. As indicated by the arrow, there is a clear indication of decrease (pulse on) and increase (pulse off) in the height of the scattering peak / intensity with time. It is accompanied by a systematic shift of the mean peak position ($Q_0$). The error bars represent the standard deviation.

To explore the changes more in detail we have modelled the data in Figure 3 by a log-normal distribution:

$$I(Q) = \frac{A}{\sigma Q \sqrt{2\pi}} \exp\left[\frac{-\ln(Q/Q_0)^2}{2\sigma^2}\right] + B \qquad (1)$$

Here, $A$, the area under the scattering curve, $Q_0$, the mean peak position of the scattering peak, $\sigma$, characterizes the standard deviation of the distribution and $B$ the incoherent background. It should be noted that, $A$, is directly related to the amplitude of the scattered intensity, which is proportional to the volume and aggregation number of the scattered micelles.[28, 29, 49]

The symbols in Figure 4 (a) presents the normalized area, $A(t)/A(t = 0\ s)$. The sonication is indicated by the shaded area. Starting the sonication (pulse-on) there is a rapid decrease of the area to 60% of the initial unperturbed value after a pulse duration of, $t_p$ = 65 s. After the sonication is switched off the initial unperturbed value is reached by an exponential increase with time with a time constant, $t_c(A)$ = 156 ± 5 s. After sonication is switched off the system reaches the initial unperturbed value for t ≥ 900 s within the statistical accuracy. The time constant $t_c$ of our exponential growth process corresponds to the time, $t$, it takes to reach, $A(t = t_c)/A(0) = 1 - 1/e$, (~ 63.2%) of its asymptotic value. This shows the plateau value is reached after, $t > 5t_c$. In case of Figure 4 (a), the initial unperturbed (plateau) is reached for t ≥ 900 s, within the statistical accuracy. The corresponding inter-particle distance, $d_0 = 2\pi/Q_0$, is plotted in Figure 4(b). Here the system also shows an exponential recovery to the equilibrium distance of 67.30 ± 0.02 Å. We obtain a time constant, $t_c(d_0)$ = 185 ± 5 s. At the end of the ultrasonic pulse, $d_0$, was reduced to 63.16 Å, about ~ 6% of the equilibrium distance. The standard deviation, $\sigma$, is only slightly perturbed. The average value $\sigma$ = 0.243 ± 0.002 is identical to that of the unperturbed micellar structure. It should be noted that a difference between the time constants, $t_c(A)$, and, $t_c(d_0)$, is expected. Here, $t_c(A)$, represents the micellar reformation time associated with the area of the scattering intensity. For similar peak width ($\sigma$) and shape it is proportional to the volume ($V_{micelle}$) and, $N_{agg}$, of the individual scattered micelles, $A \sim V_{micelle} N_{agg}$.[28, 29, 49, 50] Whereas, $t_c(d_0)$, is associated with the recovery time of the inter-micellar distance and is related to their interaction potential.[28, 29] In fact, the dynamics associated with the individual micelle is faster than that of the collective micelles ($t_c(A) < t_c(d_0)$).

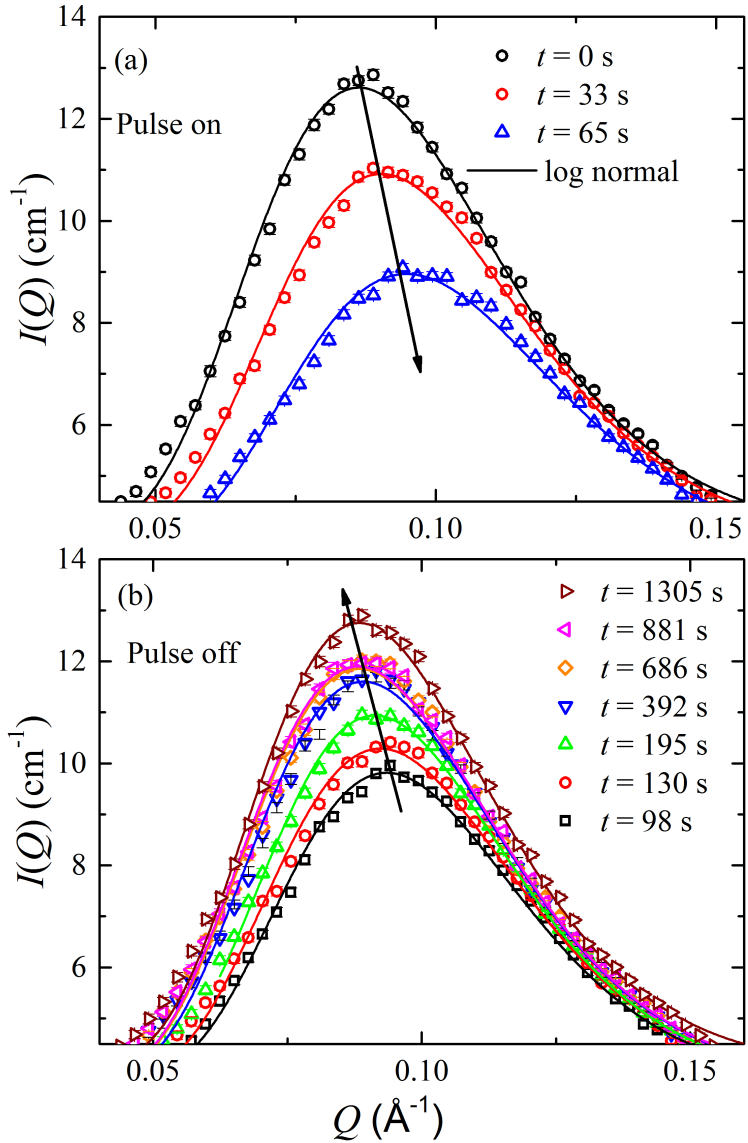

Fig.3. (1 column, color) *Time resolved SANS data for SDS micelles in D$_2$O, $\phi_w$ = 5% at 25°C. After the ultrasonic pulse was turned (a) on and (b) off. The full lines represent the parametrization by Eq. 1. Each scattering curve represents sample scattering with 30s collection time.*

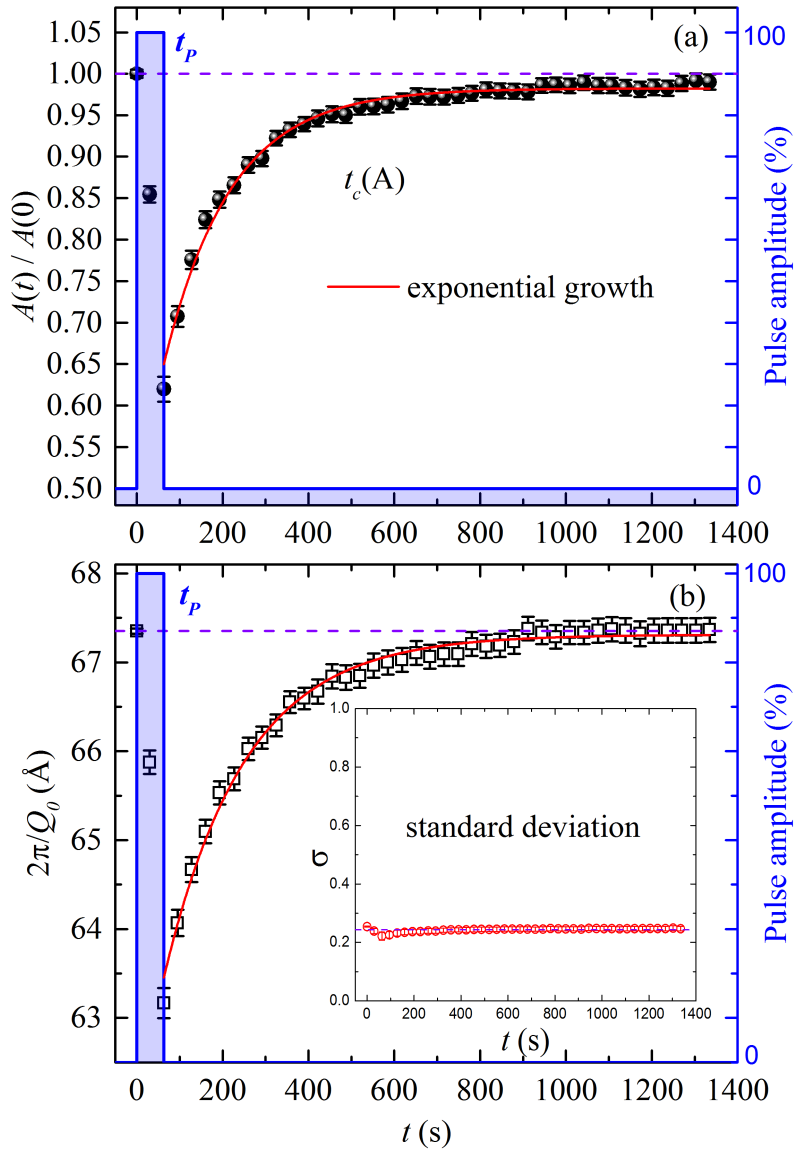

Fig.4. (1 column, color) *(a) Normalized area, A(t)/A(0), (b) inter-particle distance, $2\pi/Q_0$, from the mean peak position (equation 1) and (Inset) standard deviation, $\sigma$. The (blue) shaded area illustrates the sonication, with a duration of $t_p$ = 65 s. The solid lines represent fits by*

*exponential equations, cf. text. The dashed lines represent the unperturbed values. The dotted line represents the average value of the standard deviation, σ = 0.243 ± 0.002.*

To test the reproducibility of the experiments and/or a possible degradation of the sample, we performed multiple test experiments. In each, we sonicated the sample for $t_p$ = 65 s (pulse on) followed by a waiting time of $t_{off}$ = 1305 s (pulse off). Figure 5 represent the area obtained from equation 1 for five different on-off cycles. The first data point in each cycle mark the initial state (pulse off), followed by two data points in pulse on state that shows a rapid decay of the scattering intensity and area. It is followed by the off state waiting time that represents an exponential growth to the initial state. The error bars in the data represent the standard deviation obtained from the fitting.

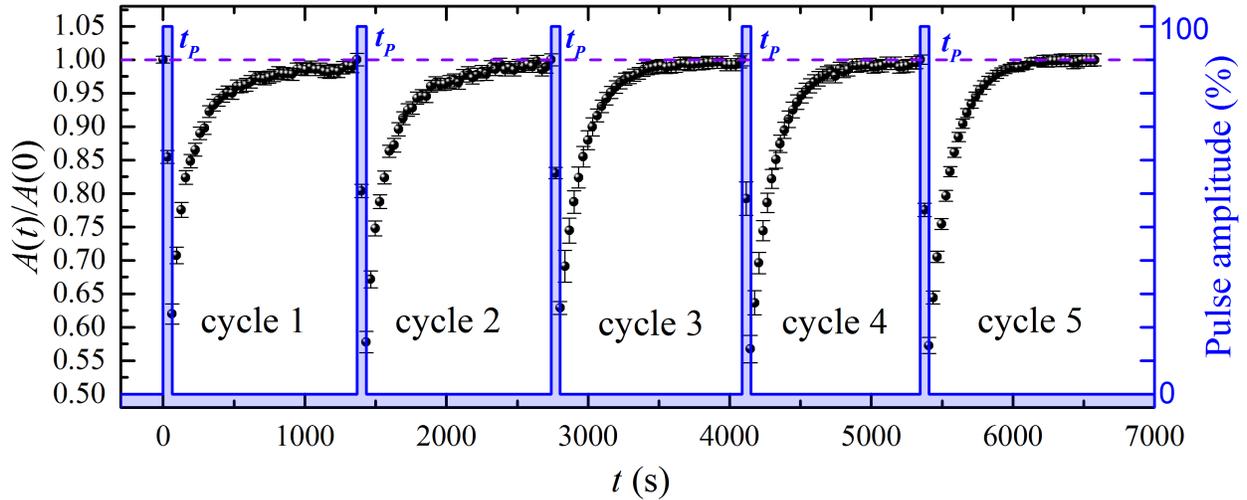

Fig.5. (2 column, color) *(a) Normalized area, A(t)/A(0), as a function of total time obtained from the SANS time resolved scattering data (Figure 3) from equation 1 with an applied fixed pulse amplitude and fixed pulsed duration $t_p$ = 65 s for five different on-off cycles.*

In order to calculate the statistically averaged response over a time period of 110 minutes, the normalized area, *A(t)/A(0)*, from Figure 5 is plotted as a function of time within each cycle (Figure 6(a)). There is no indication for systematic changes as evident from the Figure 6(a). Such statistical nature of the exponential growth opens the opportunity to accumulate the intensity of several cycles to increase the signal-to-noise ratio or to reduce the acquisition time. We observe the same behavior as described earlier for Figure 3. The mean response of micellar recovery following ultrasonic irradiation is obtained in Figure 6(b) as the normalized mean area, $\langle A(t)/A(0) \rangle$, from five different on-off cycles. The data is modelled for an exponential growth, yielding a mean time constant, $\langle t_c(A) \rangle$ = 169 ± 2 s. A similar analysis for the mean inter-particle distance yields a time constant, $\langle t_c(d_0) \rangle$ = 202 ± 2 s and a mean standard deviation, $\langle \sigma \rangle$ = 0.243 ± 0.0002.

To understand the micellar disintegration during sonication we have performed similar analysis for three consecutive on-off cycles with an ultrasonic pulse duration, $t_p$ = 112 s. The corresponding SANS diffraction pattern follow a similar trend like Figure 3(a). It was analyzed using equation 1 and the mean value of the normalized area, $\langle A(t)/A(0) \rangle$, and the inter-particle distances, $\langle d_0 \rangle = \langle 2\pi/Q_0 \rangle$, are plotted in Figures 7(a) and 7(b), respectively. They exhibit exponential decay with mean decay constants, $\langle t_c(A) \rangle$ = 48.6 ± 3 s and $\langle t_{decay}(d_0) \rangle$ = 37 ± 3 s, as indicated by the solid lines. Inset in Figures 7(b) illustrates the corresponding mean standard deviation, $\langle \sigma \, (decay) \rangle$ = 0.23 ± 0.002.

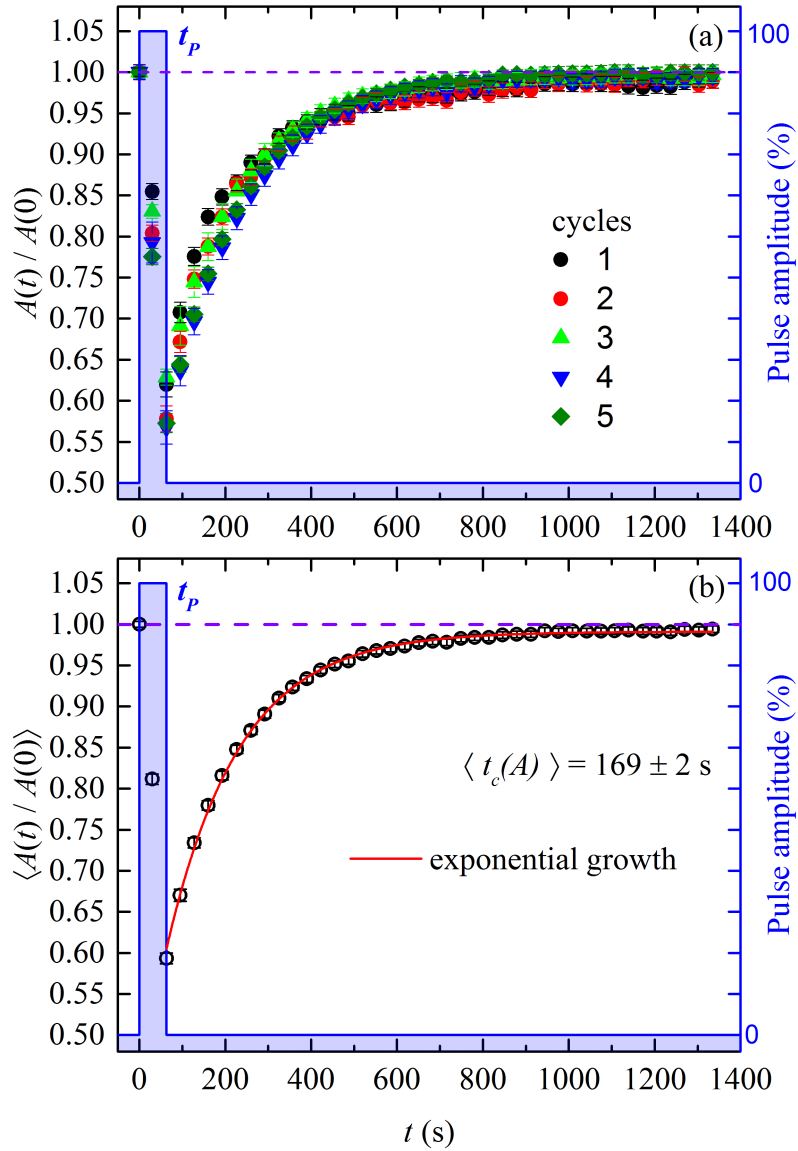

Fig.6. (1 column, color) *(a) Normalized area, A(t)/A(0), as a function of time, reset after every cycle (b) Normalized mean area, ⟨A(t)/A(0)⟩, for five on-off cycles as a function of time. Solid line represents an exponential growth with a time constant, ⟨$t_c(A)$⟩.*

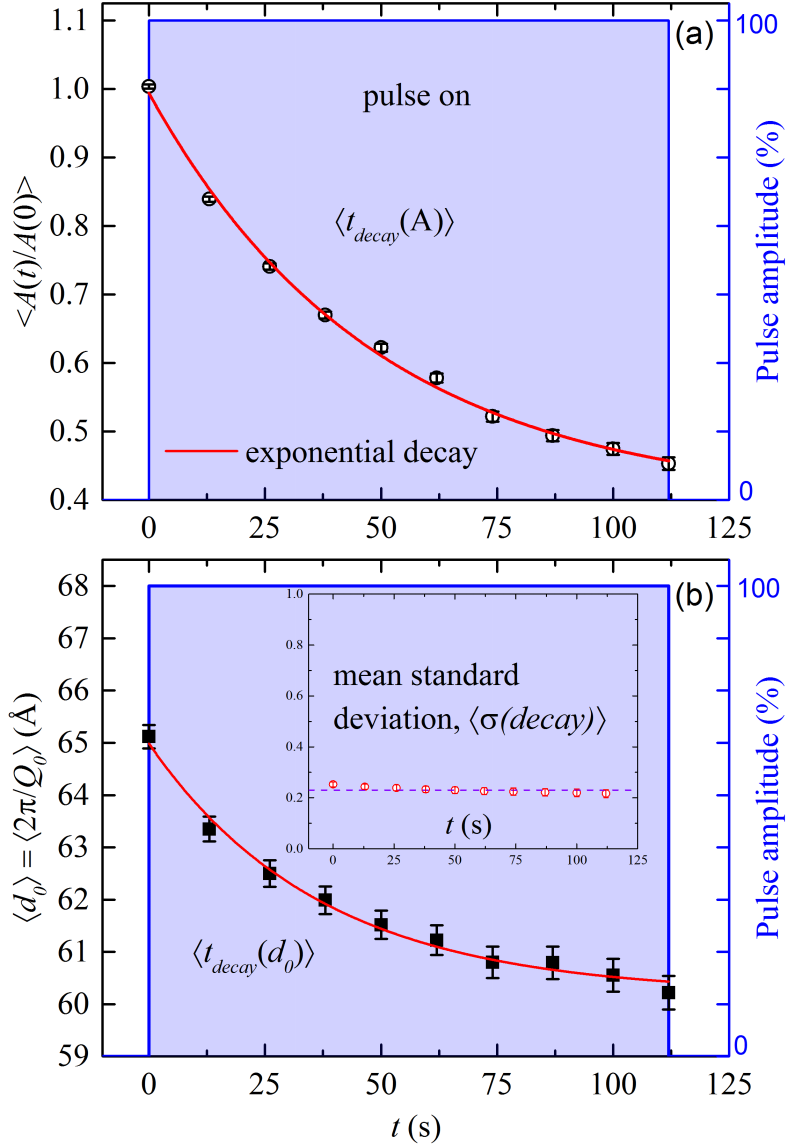

Fig.7. (1 column, color) *During sonication for three consecutive on-off cycles, (a) mean normalized area, ⟨A(t)/A(0)⟩, and (b) mean inter-particle distance, ⟨$d_0$⟩, from fitting the corresponding SANS diffraction (equation 1) as a function of time, reset after every cycle. Solid line represents an exponential decay with decay time constants, ⟨$t_{decay}(A)$⟩ and ⟨$t_{decay}(d_0)$⟩. Inset, represents the mean standard deviation, ⟨σ (decay)⟩.*

It should be noted that the set of parameters immediately after the ultrasonic pulse, ⟨$t_c(A)$⟩, ⟨$t_c(d_0)$⟩, and ⟨σ⟩, reflects the statistical average micellar reformation time, recovery of the inter-particle distances and the associated polydispersity, respectively. During the ultrasonic pulse a similar set of parameters, $\langle t_{decay}(A) \rangle$, $\langle t_{decay}(d_0) \rangle$, and $\langle \sigma(decay) \rangle$, reflects statistical average perturbed decay time depicting micellar disintegration, decaying of the inter-particle distances from its equilibrium value and the associated polydispersity, respectively. A possible explanation for such a time constant can be associated with the formation of the micelles after an initial breakdown of the entire micelles during sonication.[57, 58] It should be noted that such a process is associated with the change in both size and aggregation number of the micelles. A detailed further investigation of such phenomena is required.

**Alkyl-PEO micelles**

Figure 8 illustrates the SANS diffraction data for commercially available 18-alkyl-poly(ethylene oxide) ($C_{18}$-PEO5) in deuterated water at a volume fraction, ϕ= 0.5%. Here the 1-D scattering data is obtained by summing 10 scattering curves with each measured for 12 s during the ultrasound pulse on state. During the pulse off state, we performed summation over 20 separate scattering curves with each measured for 21 s. The solid line represents the calculated scattering intensity using a frozen micelle form factor[28] using a micellar radius, $R_m$ = 8.6 nm, taken from the literature.[54] We did not see a difference in the diffraction pattern.

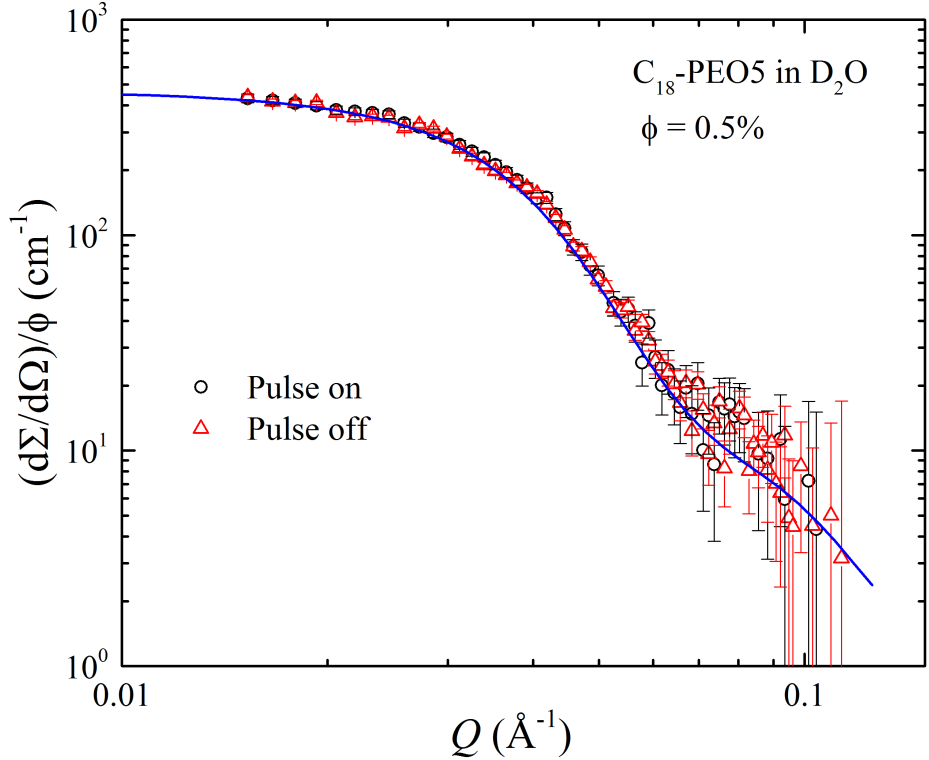

Fig.8. (2 column, color) *SANS diffraction for 0.5%, surfactant $C_{18}$-PEO5 in $D_2O$, with and without being irradiated by an ultrasonic pulse. The solid line is calculated based on modified micellar form factor.*[28]

## VI. SUMMARY

In summary, we have designed and implemented a new ultrasonic transducer sample cell for neutron scattering, especially adapted for the standard SANS sample holder and temperature control. The cell enables us to perform a unique, first of its kind time resolved *in situ* SANS experiment by periodically switching on and off the high intensity ultrasonic pulse. Using our cell, we successfully demonstrated test measurements to determine the time constant associated with SDS micellar self-assembly followed by ultrasound induced disintegration at a fixed temperature. We have also presented a weak scattering case for frozen alkyl-PEO micelles where we do not see any ultrasound induced structural change. Additional advantage of the cell is a very low incoherent background that will allow us to gain structural information over a broader $Q$- range.

## ACKNOWLEDGEMENTS

We acknowledge the support of Louisiana Consortium for Neutron Scattering (LaCNS). The neutron scattering work is supported by the U.S. Department of Energy (DoE) under EPSCoR Grant No. DE-SC0012432 with additional support from the Louisiana Board of Regents. We would like to acknowledge Paul Butler and Cedric Gagnon from the National Institute of Standards and Technology Center for Neutron Research (NIST-NCNR) for offering SANS time, helpful discussions and technical assistance. We would acknowledge Christopher Van Leeuwen, department of Chemistry, Louisiana State University, for carefully proofreading the manuscript. The authors acknowledge the support of the National Institute of Standards and Technology (NIST), U.S. Department of Commerce, in providing the neutron research facilities used in this work. Access to NGB30SANS was provided by the Center for High Resolution Neutron Scattering, a partnership between the NIST and the National Science Foundation under Agreement No. DMR-1508249.

[1] G. K. Lewis, Jr. and W. L. Olbricht Review of Scientific Instruments **79,** (2008).
[2] F. Lefebvre, J. Petit, G. Nassar, P. Debreyne, G. Delaplace and B. Nongaillard Review of Scientific Instruments **84,** (2013).
[3] B. Gerold, S. Kotopoulis, C. McDougall, D. McGloin, M. Postema and P. Prentice Review of Scientific Instruments **82,** (2011).
[4] Y. C. Wang and R. S. Lakes Review of Scientific Instruments **74,** (2003).
[5] T. G. Hertz, S. O. Dymling, K. Lindström and H. W. Persson Review of Scientific Instruments **62,** (1991).
[6] A. Sun, X. Bai and B. F. Ju Review of Scientific Instruments **86,** (2015).
[7] S. Baer, M. A. Andrade, C. Esen, J. C. Adamowski, G. Schweiger and A. Ostendorf Review of Scientific Instruments **82,** (2011).
[8] E. C. Everbach The Journal of the Acoustical Society of America **118,** (2005).

[9]R. E. Challis, M. J. W. Povey, M. L. Mather and A. K. Holmes Reports on Progress in Physics **68,** (2005).
[10]C. Huang, G. W. Auner, H. J. Caulfield and J. D. G. Rather Acoustics Research Letters Online **6,** (2005).
[11]J. Han, G. Fei, G. Li and H. Xia Macromolecular Chemistry and Physics **214,** (2013).
[12]E. C. Unger, T. Porter, W. Culp, R. Labell, T. Matsunaga and R. Zutshi Adv Drug Deliv Rev **56,** (2004).
[13]A. Azagury, L. Khoury, G. Enden and J. Kost Adv Drug Deliv Rev **72,** (2014).
[14]A. K. Wood and C. M. Sehgal Ultrasound Med Biol **41,** (2015).
[15]B. Hammouda, D. Ho and S. Kline Macromolecules **35,** (2002).
[16]A. A. Lefebvre, J. H. Lee, N. P. Balsara and B. Hammouda Journal of Polymer Science Part B: Polymer Physics **38,** (2000).
[17]S. L. Pesek, X. Li, B. Hammouda, K. Hong and R. Verduzco Macromolecules **46,** (2013).
[18]D. Richter, D. Schneiders, M. Monkenbusch, L. Willner, L. J. Fetters, J. S. Huang, M. Lin, K. Mortensen and B. Farago Macromolecules **30,** (1997).
[19]J. Stellbrink, J. Allgaier, L. Willner, D. Richter, T. Slawecki and L. J. Fetters Polymer **43,** (2002).
[20]T. Glomann, A. Hamm, J. Allgaier, E. G. Hubner, A. Radulescu, B. Farago and G. J. Schneider Soft Matter **9,** (2013).
[21]K. Nusser, T. Mosbauer, G. J. Schneider, K. Brandt, G. Weidemann, J. Goebbels, H. Riesemeier and D. Göritz Journal of Non-Crystalline Solids **358,** (2012).
[22]K. Nusser, S. Neueder, G. J. Schneider, M. Meyer, W. Pyckhout-Hintzen, L. Willner, A. Radulescu and D. Richter Macromolecules **43,** (2010).
[23]K. Nusser, G. J. Schneider, W. Pyckhout-Hintzen and D. Richter Macromolecules **44,** (2011).
[24]G. J. Schneider, W. Hengl, K. Brandt, S. V. Roth, R. Schuster and D. Göritz Journal of Applied Crystallography **45,** (2012).
[25]G. J. Schneider, K. Nusser, S. Neueder, M. Brodeck, L. Willner, B. Farago, O. Holderer, W. J. Briels and D. Richter Soft Matter **9,** (2013).
[26]G. J. Schneider, K. Nusser, L. Willner, P. Falus and D. Richter Macromolecules **44,** (2011).
[27]G. J. Schneider, V. Vollnhals, K. Brandt, S. V. Roth and D. Goritz J Chem Phys **133,** (2010).
[28]S. Gupta, M. Camargo, J. Stellbrink, J. Allgaier, A. Radulescu, P. Lindner, E. Zaccarelli, C. N. Likos and D. Richter Nanoscale **7,** (2015).
[29]M. Laurati, J. Stellbrink, R. Lund, L. Willner, D. Richter and E. Zaccarelli Phys Rev Lett **94,** (2005).
[30]R. Lund, V. Pipich, L. Willner, A. Radulescu, J. Colmenero and D. Richter Soft Matter **7,** (2011).
[31]S. Xuan, S. Gupta, X. Li, M. Bleuel, G. J. Schneider and D. Zhang Biomacromolecules **18,** (2017).
[32]S. Gupta, R. Biehl, C. Sill, J. Allgaier, M. Sharp, M. Ohl and D. Richter Macromolecules **49,** (2016).
[33]A. Shukla, E. Mylonas, E. Di Cola, S. Finet, P. Timmins, T. Narayanan and D. I. Svergun Proc Natl Acad Sci U S A **105,** (2008).
[34]S. Gupta, N. Arend, P. Lunkenheimer, A. Loidl, L. Stingaciu, N. Jalarvo, E. Mamontov and M. Ohl Eur Phys J E Soft Matter **38,** (2015).
[35]S. Gupta, J. K. H. Fischer, P. Lunkenheimer, A. Loidl, E. Novak, N. Jalarvo and M. Ohl Scientific Reports **6,** (2016).
[36]S. Gupta, E. Mamontov, N. Jalarvo, L. Stingaciu and M. Ohl Eur Phys J E Soft Matter **39,** (2016).
[37]J. Wuttke Review of Scientific Instruments **83,** (2012).
[38]F. Kaneko, N. Seto, S. Sato, A. Radulescu, M. M. Schiavone, J. Allgaier and K. Ute Journal of Physics: Conference Series **746,** (2016).
[39]F. Kaneko, N. Seto, S. Sato, A. Radulescu, M. M. Schiavone, J. Allgaier and K. Ute J Appl Crystallogr **49,** (2016).
[40]W. Pyckhout-Hintzen, S. Westermann, A. Wischnewski, M. Monkenbusch, D. Richter, E. Straube, B. Farago and P. Lindner Physical Review Letters **110,** (2013).
[41]G. J. Schneider and D. Goritz J Chem Phys **133,** (2010).
[42]J. Kalus, G. Neubauer and U. Schmelzer Review of Scientific Instruments **61,** (1990).
[43]T. C. B. McLeish Advances in Physics **51,** (2002).

[44]M. Rubinstein and R. H. Colby, *Polymer Physics* (2007).
[45]S. Gupta, S. K. Kundu, J. Stellbrink, L. Willner, J. Allgaier and D. Richter J Phys Condens Matter **24,** (2012).
[46]C. R. Lopez-Barron, L. Porcar, A. P. Eberle and N. J. Wagner Phys Rev Lett **108,** (2012).
[47]M. A. Calabrese, N. J. Wagner and S. A. Rogers Soft Matter **12,** (2016).
[48]S. Gupta, J. Stellbrink, E. Zaccarelli, C. N. Likos, M. Camargo, P. Holmqvist, J. Allgaier, L. Willner and D. Richter Phys Rev Lett **115,** (2015).
[49]R. Lund, L. Willner, J. Stellbrink, P. Lindner and D. Richter Phys Rev Lett **96,** (2006).
[50]R. Lund, L. Willner, V. Pipich, I. Grillo, P. Lindner, J. Colmenero and D. Richter Macromolecules **44,** (2011).
[51]M. Bergström and J. Skov Pedersen Physical Chemistry Chemical Physics **1,** (1999).
[52]B. Hammouda J Res Natl Inst Stand Technol **118,** (2013).
[53]C. Sommer, J. S. Pedersen and V. M. Garamus Langmuir **21,** (2005).
[54]T. Zinn, L. Willner and R. Lund Phys Rev Lett **113,** (2014).
[55]C. J. Glinka, J. G. Barker, B. Hammouda, S. Krueger, J. J. Moyer and W. J. Orts Journal of Applied Crystallography **31,** (1998).
[56]S. R. Kline Journal of Applied Crystallography **39,** (2006).
[57]E. A. G. Aniansson and S. N. Wall The Journal of Physical Chemistry **78,** (1974).
[58]J. Lang, C. Tondre, R. Zana, R. Bauer, H. Hoffmann and W. Ulbricht The Journal of Physical Chemistry **79,** (1975).